\documentstyle[aaspp4,epsfig]{article}

\newcommand{\rmln}{{\rm ln}}

\newcommand{\calT}{{\cal T}}

\newcommand{\ba}{\begin{array}}
\newcommand{\ea}{\end{array}}
\newcommand{\beq}{\begin{equation}}
\newcommand{\eeq}{\end{equation}}

\begin{document}

\title{PLASMA TURBULENCE AND STOCHASTIC ACCELERATION IN SOLAR FLARES}
\author{Vah\'e Petrosian\altaffilmark{1},}
\affil{Center for Space Science and Astrophysics, Stanford University,
Stanford,
CA  94305}

\altaffiltext{1}{also Departments of Physics and  Applied Physics}


\begin{abstract}

Observational aspects of solar flares relevant to the acceleration process of
electrons and protons are reviewed and it is shown that most of these
observations can be explained by the interaction with flare plasma of a power
law energy distribution of electrons (and protons) that are injected at the top
of a flaring loop, in the so-called thick target model.  Some new observations
that do not agree with this model are described and it is shown that these can
be explained most naturally if most of the energy released by the reconnection
process goes first into the generation of plasma turbulence, which accelerates,
scatters and traps the ambient electrons near the top of the loop
stochastically.  The resultant bremsstrahlung photon spectral and spatial
distributions agree with the new observations.  This model is also justified by
some theoretical arguments.  Results from numerical evaluation of the spectra of
the accelerated electrons and their bremsstrahlung emission are compared with
observations and shown how one can constrain the model parameters describing the
flare plasma and the spectrum and the energy density of the turbulence.
\end{abstract}

\section{INTRODUCTION}

Because of the proximity of the Sun, solar flares have been detected in
multitude of ways and there is considerable body of relatively detailed
observations, a fact uncommon in other similar astrophysical sources and
situations.  Consequently, solar flares provide an excellent labaratory for
testing ideas and models for all aspects of high energy astrophysical phenomena
such as enenrgy generation and release, acceleration of particles and radiation
processes.  In this paper I describe a somewhat new paradigm for particle
acceleration in solar flares where plasma waves or turbulence play a much more
dominant role than has been attributed to them in the past.  In \S 2 I present a
brief review of some of the observational features of solar flares most relevant
to the acceleration process, and a short outline of current ideas on production
(and channels of release) of flare radiations.  Then in \S 3 I describe some new
observations that cannot easily be accounted for by the standard model, but can
come about naturally in a model where particles are accelerated stochastically
by plasma turbulence.  Here I also give some theoretical justifications for such
a model.  In \S 4 I develop a simplified model of acceleration based on this
idea and in \S 5 compare the predictions of this model with the new
observations.  In \S 6 I give a brief summary.

\section{A BRIEF REVIEW}

Flare radiation has been observed throughout the whole range of electromagnetic
spectrum from long wavelength radio waves to gamma-rays up to hundreds of Mev.
At radio wavelengths there are the various types (II, III, IV, V) of long
wavelength emissions, but most relevant to the acceleration process is the
continuum emission in the microwave-submillimeter range.  Notable in the optical
UV range are the H$_\alpha$ and white light emissions.  In the soft X-ray range
($<$ 10 keV), there is the thermal continuum emission and line emission by
highly ionized heavy elements.  Above 10 keV, in the hard X-ray (10 keV to 1
MeV) and gamma-ray ($>$ 1 MeV) range, there is a non-thermal continuum radiation
resulting from bremsstrahlung of the accelerated electrons, and in the 1 to 7
MeV range there is the gamma-ray line emission produced by accelerated protons
and other ions.  In addition to the electromagnetic radiation, there are other
observations of flare phenomenon, among which the one most relevant to the
acceleration process is the direct observation near the earth of accelerated
electrons, protons, neutrons and other nuclei.

These radiations can be divided into two categories.  The microwave, hard X-ray
and all gamma-ray radiations are non-thermal radiation produced directly by the
accelerated particles and constitute what is called the {\it impulsive phase} of
the flares.  These radiations are highly correlated and often show almost
identical temporal evolution.  The rest of the radiations are associated with
what is called the {\it gradual phase} of the flare and are manifestations of
flare plasma energized by the primary energy release process.  These secondary
radiations evolve at a rate that is approximately proportional to the integral
of the impulsive time profiles, achieve their maximum at the end of the
impulsive phase and then decay gradually.

Initially when only thermal radiation was observed, it was believed that most of
the released energy goes directly into heating.  But since the discovery of the
non-thermal microwave and hard X-ray radiation it is believed that a large
fraction, perhaps all, of the flare energy goes into acceleration of particles,
mostly electrons with a power law distribution, $f(E)=\kappa E^{-\delta}$.
These electrons produce X-rays through collisions with the ambient protons and
ions as they travel down the flaring loop.  Some of the radiation comes from the
loop but the bulk of it is emitted in the high density regions below the
transition layer.  The total bremsstrahlung yield of X-rays with energy $k \geq
E_0$ (in units of $m_ec^2$) is $Y ={{2\pi \alpha} \over {{\rmln} \Lambda}} {E_0
{\delta - 1} \over {\delta^2(\delta - 2)}}$, where $\alpha$ is the fine
structure constant and ${\rm ln} \Lambda \simeq 20$ is the Coulomb logarithm.
For typical values of $E_0 = 20$ keV and $\delta = 4$, $Y \leq 10^{-5}$.  This
means that almost all of the accelerated electron energy goes into heating or
evaporation of plasma in the lower atmosphere, which gives rise to the softer
X-rays and other gradual thermal radiations.  This scenario agrees with the
above-mentioned integral relation, the usual power law X-ray spectra, and the
rudimentary observations of the spatial structure (double foot point sources)
observed at hard X-rays by SMM and HINATORI spacecrafts.

The electrons as they spiral down the flaring loop (defined by the magnetic
lines of force) emit also synchrotron radiation at microwave frequencies.  The
interpretation of this emission is more complicated because of its dependence on
the magnetic field strength and geometry, and the presence of various kinds of
absorptions.  Accelerated protons and ions give rise to nuclear excitation and
other gamma-ray lines in the 1-7 MeV range and to a continuum emission around 50
MeV from decay of pions they produce.  For a review of gamma-rays from protons
and ions the reader is referred to the excellent review by Ramaty and Murphy
(1989).  We shall not discuss the microwave emissions nor describe the models
for gamma-ray line emission whose presence complicates the interpretation of the
electron bremsstrahlung emission.  To avoid this complication we will deal with
the so-called electron dominated flares where a small number of ions are
accelerated and their emission can be ignored (cf.  e.g.  Marschha\"user et al. 
1991).

\section{WHY PLASMA TURBULENCE?}

There are, however, several more recent observations which do not agree with
this scenario at least in its simplest form.  In this section we describe two
such observations and show how the presence of plasma turbulence can account for
these observations.  Then we present some theoretical arguments in favor of a
model where bulk of the energy released by the reconnection process does not go
directly into heating of the plasma or acceleration of the particles but into
generation of plasma waves or turbulence which in turn can accelerate particles
and heat the plasma (directly, or indirectly via the accelerated particles).

\subsection{Observational Motivation}

The two observations that provide evidence for the above scenario are the wide
dynamic range spectral and high spatial resolution hard X-ray observations.
There may be other evidence such as the impulsive soft X-ray observation from
foot points (Hudson et al.  1994, Petrosian 1994, 1996).

\subsubsection{Spectral Evidence}

Observed spectra over a limited range (30-500 keV) can be fitted to
bremsstrahlung spectra emitted by electrons with a power law energy spectrum.
But higher resolution spectra, especially those with a wider dynamic range (10
keV to 100 MeV), observed by the gamma-ray spectrometer (GRS) onboard SMM and by
the combined BATSE and EGRET instrumentS on Compton Gamma-Ray Observatory
(CGRO), show considerable deviation from a simple power law; as shown in Figure
1, there is spectral hardening (flatter spectra) above around few 100s of keV
and a steep cutoff above 40 MeV.  These kinds of deviations can be produced by
the action of Coulomb collisions and synchrotron losses during the {\it
transport} of the electrons from top of a loop to the foot points.  However, as
shown by Petrosian, McTiernan \& Marshha\"user (1994), the observed deviations
would require plasma densities $n$ and magnetic fields $B$, which are much
higher than those believed to be present (see Figure 1).  We conclude that these
signatures must be present in the spectra of the accelerated electrons.  As we
shall show in \S 5 the stochastic acceleration model described in the next
section can reproduce these observations with more reasonable values of the
parameters.

\subsubsection{Spatial Structure}

The second observation which provides a more direct evidence for presence of
turbulence at the top of loops is the observations by YOHKOH of loop top hard
X-ray (10-50 keV) emission (Masuda et al.  1994, Masuda 1994).  Figure 2 shows
an image of one such flare along with variations with time of the emission
intensities from the loop top, the foot points and their ratio.  Several other
flares show similar images and emission ratios.  There exists also a limited
spectral information (Masuda 1994 and Alexander \& Metcalf 1997).

In the standard thick target model, the spatial variation of the bremsstrahlung
emission depends primarily on the ralative values of the length $L$ of the loop
and the Coulomb collision mean free path, $\lambda _{\rm Coul} = \beta^2E/(4\pi
r_0^2 n {\rm ln}\Lambda)$, where $\beta =v/c$ is the electron velocity in units
of the speed of light and $r_0 \simeq 2.8 \times 10^{-13}$cm is the classical
electron radius.  In general, $L \ll \lambda_{\rm Coul}$ in the corona so that
most particles travel through the coronal portion of the loop freely but undergo
rapid collision and bremsstrahlung emission once they reach the transition
region and the chromosphere where the density increases by several orders of
magnitude within a much shorter distance.  Loop top emission is possible only if
$ \lambda_{\rm Coul} \ll L$, in which case the accelerated electrons lose most
of their energy at the top of the loop with no emission from foot points.  If $
\lambda_{\rm Coul}\simeq L$ then one would expect a uniform emission from
throughout the loop.  For further deatails see J.  Leach's Ph.D.  thesis (1984),
Fletcher (1995) and Holman (1996).

\begin{figure}[htbp]
\leavevmode
\centerline{
\psfig{file=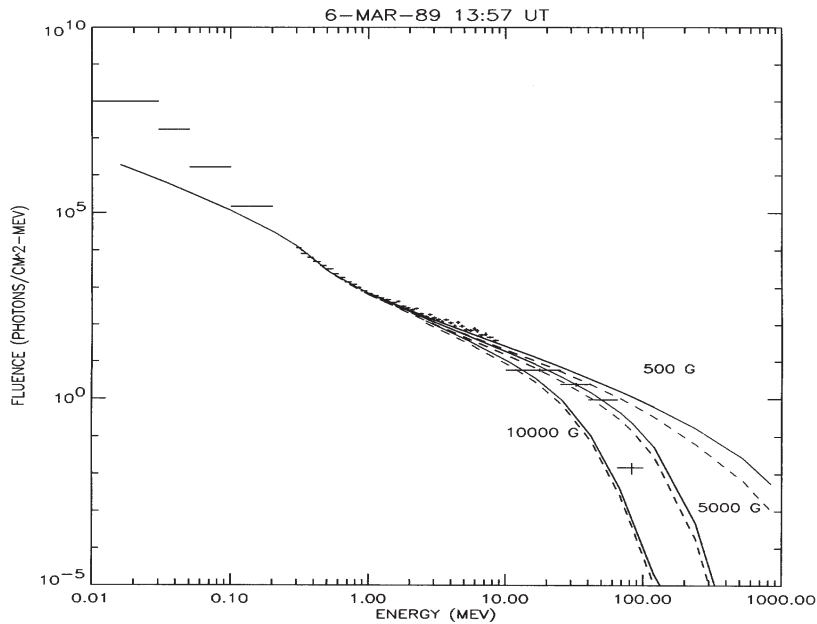,width=0.75\textwidth,height=0.6\textwidth}}
\caption{Spectrum of the first impulsive bump (13:57:29-13:58:34 UT) of the
1989 March 6 electron dominated flare. The solid lines show the whole loop 
spectrum for an
injected power law electrons (index $\delta=2.2$) adjusted to obtain best fit
in 0.3-10 MeV range. The curves are marked by the value $B_0$ at the top of
the
loop with a factor of 2.5 increase at the transition region.
The dashed lines show the effects of the optical depth.  Note
the failure  of the models at low energies and the high required
field strengths at high  energies. (From Petrosian et al.
1994.) }
\label{trans}
\end{figure}

Significant loop top emission is possible if the loop is highly inhomogeneous.
Since the emission is proportional to the ambient density a higher density at
the loop top can mimic the observed variation (Wheatland and Melrose 1995).
However, it is difficult to see how large density gradients can be maintained in
a $T \geq 10^{6}$ K plasma.  Enhancement of loop top emission can be produced
more naturally in loops with converging field geometry.  The field convergence
increases the pitch angle of the particles as they travel down the loop, and if
strong enough, can reflect them back up the loop before they reach the
transition region.  This increases the density of the accelerated electrons near
the top of the loop giving rise to a corresponding enhancement of the
bremsstrahlung emission.  Such models were also investigated by Leach (1984),
show significant increases of emission from loop tops for large convergence
ratio (see also Leach \& Petrosian 1983 and Figures 1 and 2 in Peterosian \&
Donaghy 1999; {\bf PD} for short), and by Fletcher (1996) showing increase in
the average height of the X-ray emission.

\begin{figure}[htbp]
\leavevmode
\centerline{\hspace{1.7in}
\psfig{file=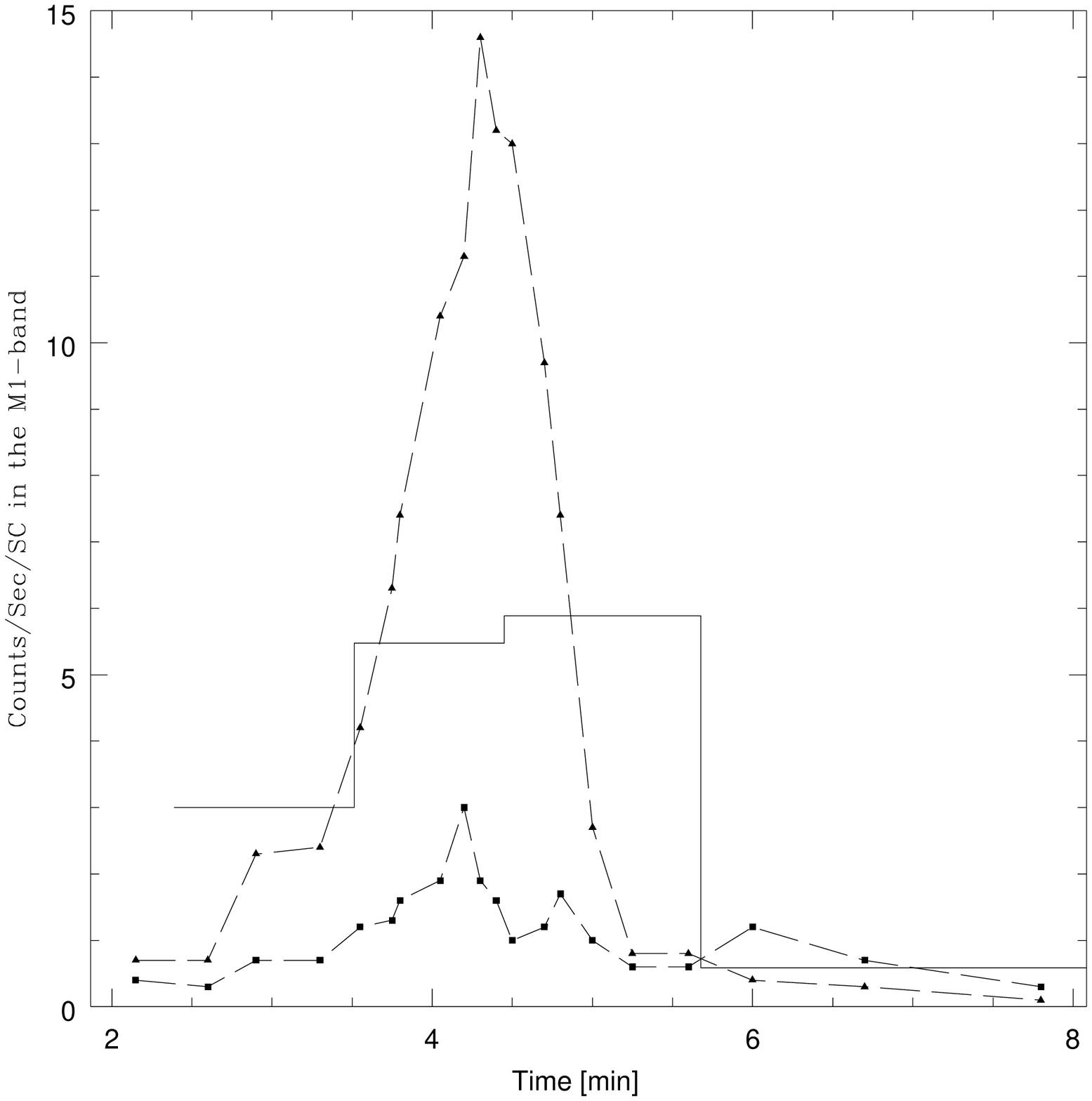,width=0.5\textwidth,height=0.5\textwidth}
\psfig{file=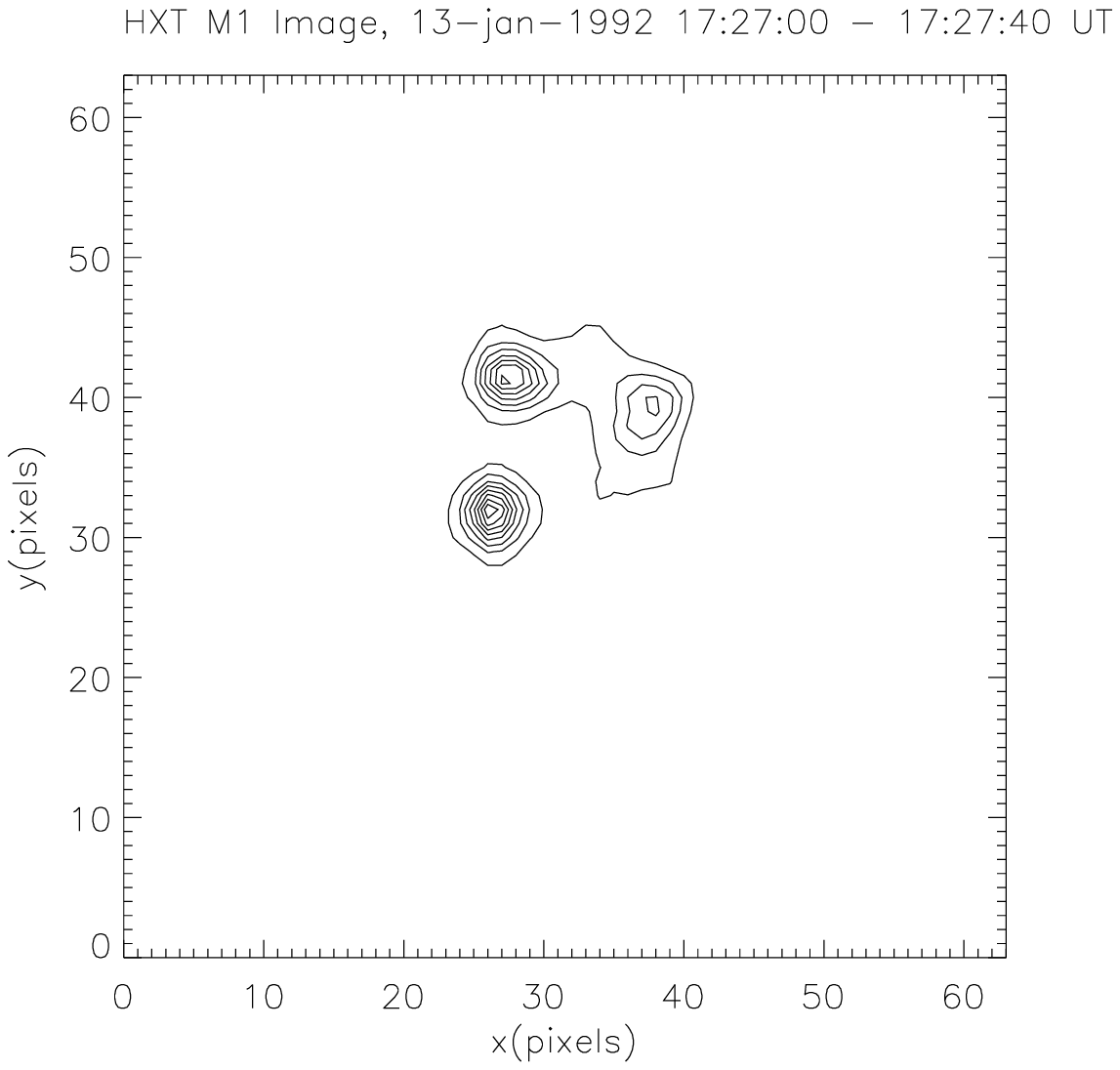,width=0.8\textwidth,height=0.5\textwidth}}
\caption{{\it Left panel} shows the intensities  of the foot points (top dashed 
curve), the loop top (lower dashed curve) and their ratio (solid histogram).  
{\it Right panel} shows the contour plots of these sources for January 13 1991 
flare. Taken from Masuda 1994.}
\label{Jan1391}
\end{figure}

Finally, trapping of electrons near the top of a loop can come about by an
enhanced pitch angle scattering with a mean free path $\lambda _{\rm scat} \ll L
\ll \lambda _{\mathrm{Coul}}$.  In this case electrons will undergo many pitch
angle scatterings and achieve an isotropic pitch angle distribution before they
escape within a time $T_{\rm esc} \simeq \frac{L}{v} \times \frac{L}{\lambda
_{\rm scat}} \gg \frac{L}{v}$.  It can be shown that the ratio of the loop top
emission to emission from other parts of the coronal portion of the loop (i.e.
the legs of the loop, which do not contain such a scattering agent but are of
comparable length) will be approximately $J_{LT} / J_{Loop} \simeq
L/\lambda_{\mathrm{scat}}$, in which case we may detect emission only from the
top of the loop and the high density foot points, but not from the legs of the
loop.  A possible scattering agent which could satisfy this condition is plasma
turbulence.  If this is the case, however, one must also include the possibility
of acceleration of the particles by the same plasma turbulence.

\subsection{Theoretical Arguments}

Various mechanisms have been proposed and are investigated for the acceleration
of particles in all kinds of astrophysical situations (see, e.g.  review
articles in the Proc.  of IAU Colloq.  142, 1994, ApJ Suppl.  90).  The three
most common mechanisms are the following.

{\bf Electric Fields} parallel to the magnetic field lines can accelerate
charged particles.  For fields greater than the Dreicer field, ${\bf E}_D =
kT/(e\lambda_{\rm Coul})$, particles of charge $e$ gain energy faster than the
mean collision time $\tau_{\rm Coul}=\lambda_{\rm Coul}/v$.  This can lead to
unstable particle distribution, which in turn can give rise to turbulence and
cause mostly heating (Boris et al.  1970, Holman 1985).  Sub-Dreicer fields, in
order to accelerate charged particles to relativistic energies, must extend over
a region $L\gg \lambda_{\rm Coul}$, which as pointed above cannot be the case
for a loop size acceleration region.  An anomalously large resistivity is one
way to get around this difficulty (Tsuneta 1985).  It seems that some other
mechanism is required for acceleration to MeV and GeV ranges (needed for
electrons and protons, respectively).

{\bf Shocks} are the most commonly considered mechanism of acceleration because
they can quickly accelerate particles to very high energies.  However this
requires the existence of some scattering agent to force repeated passage of the
particles across the shock.  The most likely agent for scattering is plasma
turbulence or plasma waves.  The rate of energy gain then is governed by the
scattering rate which in this case is proportional to the pitch angle diffusion
coefficient $D_{\mu\mu}$.  However, the turbulence needed for the scattering can
also accelerate particles stochastically (a second order Fermi process) at a
rate $D_{EE}/E^2$, where $D_{EE}$ is the energy diffusion coefficient.  At high
energies $D_{\mu\mu} \gg D_{EE}/E^2$ favoring shock acceleration.

{\bf Stochastic Acceleration} is favored for acceleration of low energy
background particles because several recent analyses by Hamilton \& Petrosian
1992 and Dung \& Petrosian 1994, and by Miller and collaborators (e.g.  Miller
\& Reames 1996, Schlickeiser \& Miller 1998) have shown that under correct
conditions plasma waves can accelerate low energy particles within the desired
times.  More importantly, as our more recent work has shown (Pryadko \&
Petrosian 1997), at low energies the above inequality is reversed, $D_{EE}/E^2
\gg D_{\mu\mu}$, so that the stochastic acceleration (rate $\propto D_{EE}/E^2$)
becomes more efficient than shock acceleration whose rate is governed by the
smaller coefficient $D_{\mu\mu}$.  {\it Thus, low energy particles are
accelerated more efficiently stochastically than by shocks.}

We can, therefore, imagine two scenarios.  In one scenario, low energy particles
are accelerated first by electric fields (or stochastically) and then to higher
energies by shocks.  In a second simpler scenario, stochastic acceleration by
turbulence does the whole process, accelerating the background particles to high
energies (cf.  Petrosian 1994).  We now describe the second scenario for flares.

\section{STOCHASTIC ACCELERATION}

Stochastic acceleration by turbulence was first proposed by Ramaty and
collaborators (see e.g.  Ramaty 1979 and Ramaty \& Murphy 1987) for solar flare
protons and ions.  As argued above this process appeares to be the most
promising process for acceleration of flare electrons as well (see also
Petrosian 1994 and 1996).  Similar arguments have been put forth by Schlickeiser
and collaborators (Schlickeiser 1989; Bech et al.  1990), and Miller and
collaborators (Miller 1991; Miller, LaRosa \& Moore 1996; Miller, Guessoum \&
Ramaty 1990).

For the purpose of comparison with the observations mentioned above we need the
energy spectrum and spatial distribution of the acceleraterd electrons.  The
exact evaluation of this requires solution of the time dependent coupled kinetic
equations for {\it waves and particles}.  This is beyond the scope of this work
and is not warranted for comparison with the existing data.  We therefore make
the following simplifying and somewhat justifiable assumptions.

We assume that the flare energizing process, presumably magnetic reconnection,
produces plasma turbulence throughout the impulsive phase at a rate faster than
the damping time of the turbulence, so that the observed variation of the
impulsive phase emissions is due to modulation of the energizing process.  This
assumption decouples the kinetic equation of waves from that of the electrons.
We also assume that the turbulence is confined to a region of size $L$ near the
top of the flaring loop where particles undergo scattering and acceleration, but
eventually escape within a time $T_{\rm esc} \simeq(1 + (L/\lambda _{\rm
scat}))\tau _{tr}$ , where $\lambda _{\rm scat}\propto 1/D_{\mu\mu}$ and $\tau
_{tr} \simeq L/v$ is the traverse time across the acceleration region.  Since
both these times are shorter than the observational integration time, we can use
the steady state equation.  And because the loop top emission requires $\lambda
_{\rm {scat}} \ll L$, we can assume isotropy of the pitch angle distribution and
evaluate the electron spectrum integrated througout the finite acceleration
site.  The kinetic equation then simplifies to 
\begin{equation} \label{KEQ}
\frac{\partial ^2}{\partial E^2} [D(E) f] - \frac{\partial}{\partial E}[(A(E) -
|\dot{E}_{L}|)f] \nonumber \\ -\frac{f}{T_{\rm esc} (E)} + Q(E)=0.
\end{equation} 
Here $D$ and $A$ are the diffusion and systematic acceleration
coefficients due to the stochastic process that are obtained from the standard
Fokker-Planck equation and are related to the coefficient $D_{EE}$ and 
\begin{equation}\label{loss} 
\dot{E}_{L} =\dot{E}_{\rm Coul} + \dot{E}_{\rm synch} = - 4\pi r_{0}^{2} c n \ln 
\Lambda /\beta - 4r_0^2B^{2}\beta^{2} \gamma^{2}/9m_ec 
\end{equation} 
describes the Coulomb and synchrotron energy loss rates (in units of $m_ec^2 $).
For the source function, $Q(E)$, we use a Maxwellian distribution with
temperature, $kT = 1.5$ keV.

Following our earlier formalism (Hamilton \& Petrosian 1992, Park, Petrosian \&
Schwartz 1997, {\bf PPS} for short) we use the following parametric forms for 
the
diffusion coefficients.
\beq
\label{coeffs}
D(E)={\cal D} \beta (\gamma \beta)^{q'} ,  \,\,\,
A(E)={\cal D} (q' +2)(\gamma \beta)^{q' -1}, \,\,\,
T_{\rm esc}(E)={\cal T}_{\rm esc}(\gamma \beta)^s/\beta + L/(\beta c
\sqrt{2}).
\eeq
Here $q' $ is related to $q$, the spectral index of the (assumed) power law
distribution for the plasma wave vectors.  The above forms are more general than
what one obtains from a single type of plasma waves, say the whistler waves
(where ${\cal D}$ and ${\cal T}_{\rm esc}$ are related and $s=q' = q - 2$).
They can accomodate other scattering processes, such as the hard sphere model
($q'=2, s=0$, see Ramaty 1979) or more general turbulence.  For a more general
spectrum of turbulence that includes all possible waves, more accurate numerical
values are obtained by Dung \& Petrosian (1994) and Pryadko \& Petrosian (1997,
1999).  For a more complete analysis one should use these numerical
results.  The existing data do not warrant considerations of such details.
However, we shall choose the parameters in the above expressions so that the
relevant parameters qualitatively behave like the ones from these numerical
results.  As shown in these papers, the rate of acceleration ${\cal D}$ and
escape time ${\calT}_{\rm esc}$ are proportional to the electron gyrofrequency,
$\Omega_e$, and the level of turbulence $f_{turb}=8\pi {\cal E}_{tot}/B^2$,
where ${\cal E}_{tot}$ is the total energy density of turbulence.

We solve equation (\ref{KEQ}) numerically for the spectrum of electrons at the
acceleration site,  $f_{AS}(E)$, and then evaluate the ({\it thin target})
bremsstrahlung spectrum, number of photons as a function of photon energy $k$
(in units of $m_ec^2$), emitted by these electrons:
\begin{equation}
\label{LoopTop}
  J_{AS}(k) = V \int_{k}^{\infty} dE f_{AS}(E) \beta c
    n_{AS} \frac{d\sigma}{dk}(E, k),
\end{equation}
Here $V$ and $n_{AS}$ are the volume and the background density in the
acceleration site, $d\sigma/dk$, given by Koch \& Motz (1959, eq.  [3BN]), is
the relativistically correct, angle integrated, bremsstrahlung cross section.
The electrons escaping the acceleration site travel to the foot points and
maintain a nearly isotropic pitch angle distribution in the downward direction
with a spectral flux equal to $F_{\rm esc}=Lf_{AS}/T_{\rm esc}$ and emit
bremsstrahlung at the foot points.  As is well known (see, e.g.  Petrosian 
1973),
the effective spectrum of the electrons (the so-called cooling spectrum) in the
{\it thick target} foot point sources is
\beq
\label{fthick} f_{\rm thick}(E)= - { 1 \over {\dot E}_L}
\int^\infty_E{f_{AS}(E)
\over T_{\rm esc}}dE,
\eeq
so that the photon spectrum at the foot points, $J_{FP}(k)$, is obtained from
equation (\ref{LoopTop}) by replacing the above spectrum for $f_{AS}$.

For a steep power law electron spectrum, ($f(E) = \kappa E^{-\delta}, \delta
\gg 1$), and for a relatively slow variation with energy $E$ of
$T_{\rm esc}$ (i.e.  small $s$), one can obtain simple analytic
expressions for the photon spectra.  For example, the ratio of
the foot point to loop top emissions can be  approximated by
the  simple expression
\begin{equation}
\label{ratio}
\frac{J_{FP}}{J_{AS}} \simeq \frac{\tau_{\mathrm{Coul}} (k)}{T_{\rm esc}
(k)} =
(4\pi r_0^2cn_{AS}{\rm ln}\Lambda{\cal T}_{\rm esc})^{-1}
\left\{ \begin{array}{ll} 2^{(2-s)/4}k^{2-s/2} & k \ll 1,
\\k^{1-s} & k \gg 1,\end{array} \right.
\end{equation}
where the Coulomb collision time is given by
\begin{equation}\label{taucoul}
\tau_{\mathrm{Coul}} (k) \equiv -\left(\frac{E}{\dot{E} _{\rm Coul}}
\right)_{E=k} =(4\pi r_{0}^{2} \ln \Lambda n_{AS} c)^{-1} \left\{ 
\begin{array}{ll}\sqrt{2} k^{3/2} & k \ll 1, \\k & k \gg 1.
\end{array} \right.
\end{equation}
Steep spectra are obtained for slow acceleration rates (low levels of
turbulence) or rapid escapes, but in the opposite case, for larger values of
${\cal D}{\cal T}$, the electron spectra become flat up to some high energy,
$E_{\rm synch}$, where sychrotron losses come in and the spectrum falls off
rapidly.  In this case the above ratio is modified by setting $k=E_{\rm synch}$.
For more details see PD.

\section{COMPARISON WITH OBSERVATIONS}

\subsection{High Resolution Total Spectra}

The high resolution and wide dynamic range spectra are available only for the
total emission and not for the loop top and foot point sources separately.  We
combine the two spectra and fit $J=J_{FP} + J_{LT}$ to the observed spectra of
electron dominated flares from GRS on SMM and BATSE-EGRET on CGRO.  Figure 3 and
Table 1 show these fits and the values of the model parameters obtained for
three different forms for the acceleration coefficients.  As evident the simple
hard sphere model fails to reproduce the spectra and can be easily ruled out.
But for a model based on the whistler waves and a more general model, we obtain
very good fits and reasonable values of the parameters.  The values for density
are somewhat larger than those assumed usullay, but the required magnetic field
and amount of turbulence, $f_{turb}$, are very reasonable.  Similar 
parameter values are obtained for two other electron dominated flares (see PPS).

\begin{figure}[htbp] \leavevmode%
\centering \centerline{
\psfig{file=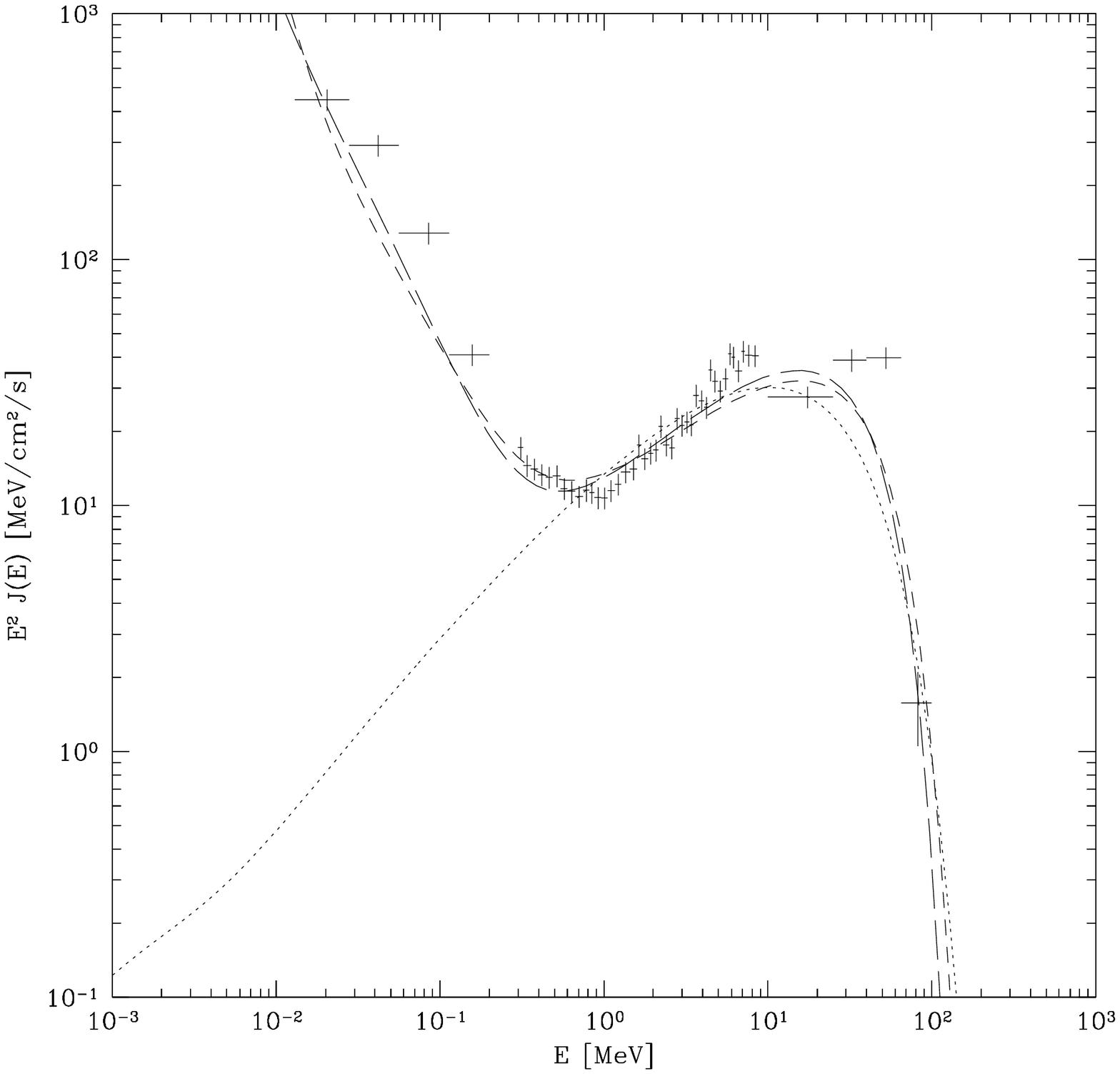,width=0.5\textwidth,height=0.5\textwidth}
\psfig{file=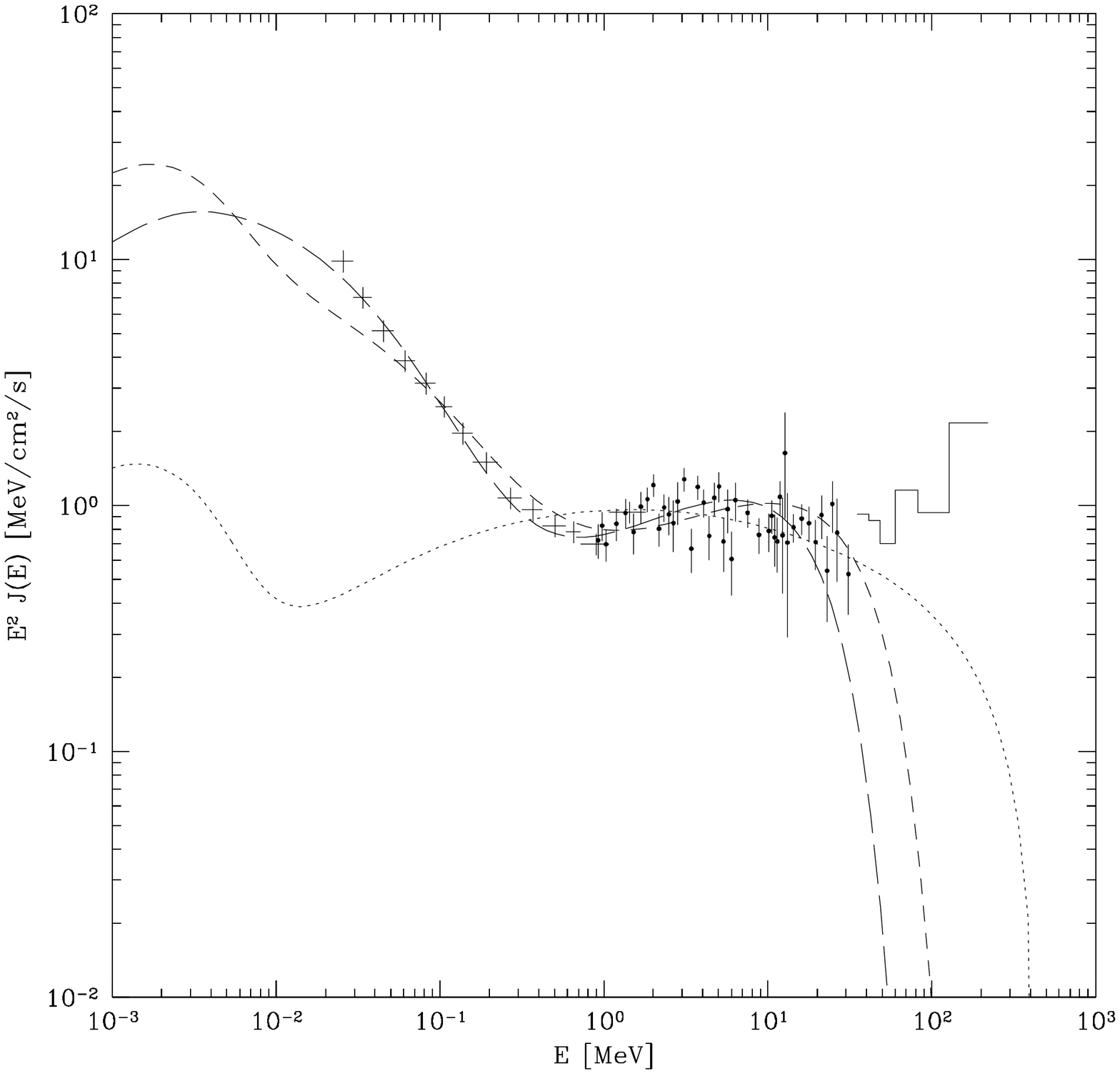,width=0.5\textwidth,height=0.5\textwidth}} 
\caption{
Comparison of the observed spectra with the total (foot point plus loop top)
photon power spectra ($E^2J(E)$):  Hard sphere (short dashed line), whistler
(medium dashed line), and the general (long dashed line).  {\it Left panel} for
the first 1989 March 6 impulsive burst (UT 13:57:29 -- 13:58:34) from GRS shown
in Fig.  1.  The fitted parameter are given in Table (1) below.  {\it Right
panel} for the BATSE (cross hair) and EGRET (solid bullet with vertical error
bars) data of the 1991 June 30 flare (UT 2:55:32 -- 2:57:11).  The solid
histogram at high energy is the EGRET 2-$\sigma$ upper limits.  The fitted
parameters for this and two other electron dominated flares are similar to those
in Table (1).  For further details see PPS.
}  
\label{spectra} 
\end{figure}

%
%

\begin{table}
\caption{
Model parameters for the first impulsive peak of the 1989 March 6 electron 
dominated flare.}
\centering
$$
\ba{clll} \hline\hline
\mbox{parameters} & \mbox{hardsphere} & \mbox{whistler} & \mbox{general} \\
\hline
Q_0 [\mbox{cm}^{-3}\mbox{s}^{-1}] & (1.4\pm0.1) \times 10^{8} & (1.7\pm 0.1) 
\times 10^{13}& (7.7\pm 5.1) \times 10^{12}\\
n [\mbox{cm}^{-3}] & (7.5\pm 0.1)\times 10^{10} & (2.9\pm 0.1)\times 10^{11} & 
(1.62\pm0.43) \times 10^{11}\\
B [\mbox{G}] & 2007\pm 29 & 500\pm 40 & 325\pm 66\\
q' & 2 & 1.45\pm 0.02 & 1.22\pm 0.09 \\
s & 0 & q' & 1.87 \pm 0.14\\
\calT [\mbox{s}] & 1.38\pm 0.03& 0.12 & 0.090\pm .006\\
L [\mbox{cm}] & 10^9& (2.8 \pm 0.4)\times 10^{9}& 10^9\\
f_{turb} & & 7\times 10^{-6}& \\
\hline
\nu & 45 & 44 & 43 \\
\chi^2/\nu & 15& 6.1& 4.9\\
\hline
\ea
$$
\label{tab:890306}
\end{table}

\subsection{High Spatial Resolution Data}

The best high spatial resolution flare data in the hard X-ray range are those
mentioned in \S 3 and shown in Figure 2. In Figure 4 we show model predictions
for the intensity ratio $J_{FP}/J_{LT}$ and the  spectal indices obtained by a
power law fit to the spectra over the short range, 20 to 50 keV, as a function
of the acceleration rate ${\cal D}$ and for several
values of the important parameters, such as density, magnetic field, and
escape time. The ranges of the observed values for these quantities
obtained by Masuda (1994) and Alexander \& Metcalf (1997) are shown by the
horizontal dotted lines.

As evident from Figure 4 some of the values of the model parameters can be
constrained by the observations.  For high ambient densitites ($n \simeq
10^{11}$cm $^{-3}$) the observed ratios agree with a wide range of the
acceleration rate except for short escape times.  But at lower densities only
smaller values of the acceleration rate are acceptable.  On the other hand, a
stronger constrain can be obtained from consideration of the spectral indices,
which require a slow acceleration rate, ${\cal D} < 0.15$s, and show a weak
dependence on the ambient density and escape time ${\cal T}_{\rm esc}$.  There
is also some dependence on the exponents $q'$ and $s$ (not shown here) and
essentially no dependence on the value of the magnetic field $B$ whose effect
come in above MeV photon energies.  The parameters from this comparison are in
agreement with those given in Table 1 obtain from spectral fittings to
different flares.  For further details the reader is referred to PD.
\begin{figure}[htbp]
\leavevmode
\centerline{
\psfig{file=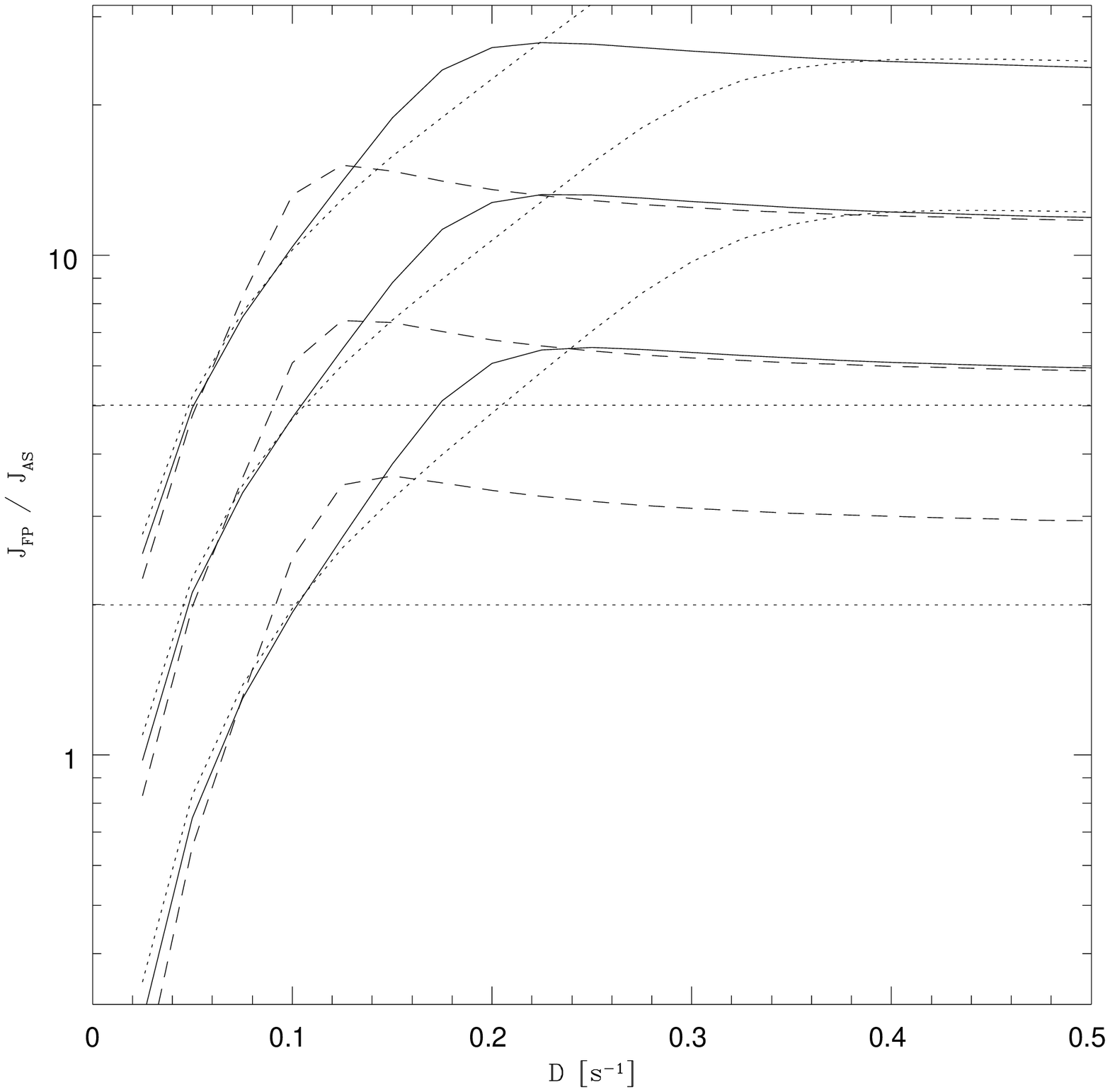,width=0.5\textwidth,height=0.5\textwidth}
\psfig{file=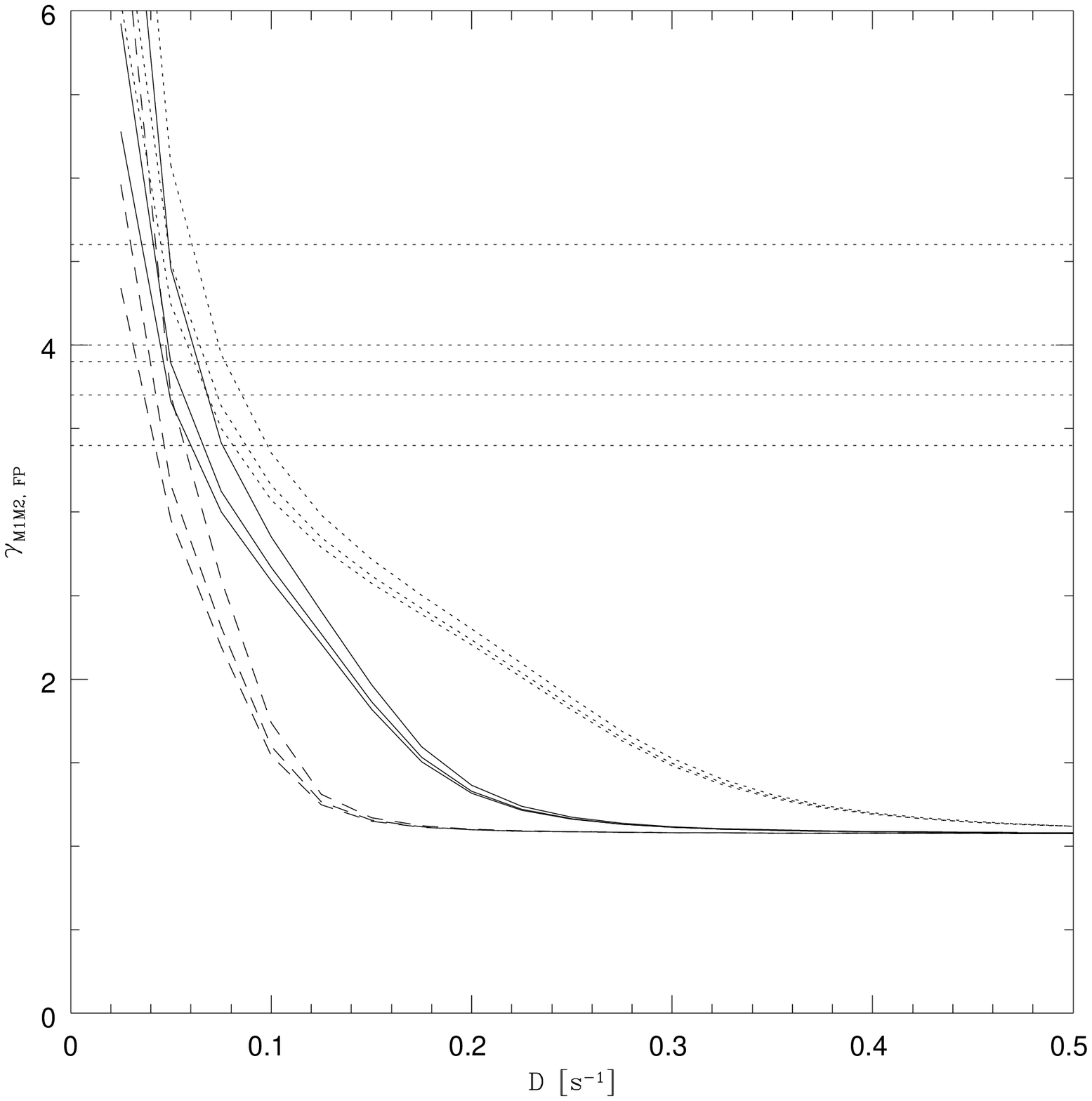,width=0.5\textwidth,height=0.5\textwidth}}
\caption{{\it Left panel}: The ratio $J_{FP} / J_{LT}$ in the 23 to 33 keV
range (the M1 band) as
a function of $ {\cal D}$ for three
escape times ${\cal T}_{\rm esc}= 0.05, 0.1$  and 0.2 s; dotted, solid and
dashed lines, respectively, and for three values of density $n/(10^{10} {\rm
cm}^{-3})=2.5, 5$ and 10; in each case from top to bottom, respectively.  The
magnetic field $B=300$G, $s=1.5$ and $q' =1.7$. The horizontal dotted lines
show the  range of the observed ratios. {\it Right panel}:
The photon number spectral index
 ($\gamma =- d{\rm ln}J(k)/d{\rm ln}k$) between channels M1 and M2
for the foot point sources. Note that  the order of the
densities here is the opposite of that in the left panel,
increasing from bottom to top. The horizontal dotted lines
show  the observed  values of this index.  From PD.
}
\label{model}
\end{figure}

\section{SUMMARY}

I have shown that some recent observations do not agree with
the predictions of the standard thick target model for the
impulsive phase of solar flares.  I have demonstrated that
higher resolution spectral observations over a wide dynamic
range and some high spatial resolution observations in the hard
X-ray regime can be explained by a modification of the standard
model where the electrons are accelerated stochastically by
plasma turbulence.  Based on models obtained in PPS and PD, I
have shown that the resultant model parameters  are
reasonable.  A more accurate determination of the validity of
the model and the range of its parameters can be obtained with
simultaneous high spectral and spatial resolution observation
of many flare that is expected during the upcoming solar maximum from
HESSI, a new  mission to be launched some time in 2000.

I would like to thank Tim Donaghy and Jim McTiernan for help with preparation of 
the figures. This work is supported in part by NASA grant NAG-5-7144-0002.

\end{document}